\RequirePackage{fix-cm}
\documentclass[twocolumn]{svjour3}          % twocolumn
\usepackage{amssymb}
\usepackage{fixltx2e}
\usepackage{graphicx}
\usepackage[english]{babel}
\usepackage{amsmath}
\usepackage{amsxtra}
\usepackage{breqn}
\usepackage{multirow}
\usepackage{algorithm}
\usepackage{algpseudocode}
\usepackage{rotating}
\usepackage{wrapfig}
\usepackage[pagewise]{lineno}
\usepackage{url}
\usepackage{natbib}
\smartqed  % flush right qed marks, e.g. at end of proof
\usepackage{mathptmx}      % use Times fonts if available on your TeX system
\usepackage{latexsym}
\journalname{  \centering \includegraphics[width=0.98\textwidth]{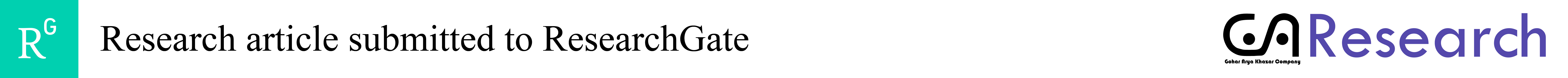}}
\begin{document}

\title{Academic Presenter:}
\subtitle{a New Storytelling Presentation Software for Academic Purposes}

\titlerunning{Academic Presenter}        % if too long for running head
\author{}
\author{Bihter Av\c{s}ar         \and
        Danial Esmaeili Aliabadi \and %etc.
        Edris Esmaeili Aliabadi  \and
        Reza Yousefnezhad
}


\institute{}
\institute{Bihter Av\c{s}ar \at
              Sabanci University, Faculty of Engineering and Natural Science, Istanbul, Turkey. \\
              \email{Bihteravsar@sabanciuniv.edu}
           \and
           Danial Esmaeili Aliabadi \at
              Sabanci University, Faculty of Engineering and Natural Science, Istanbul, Turkey. \\
              \email{Danialesm@sabanciuniv.edu}
           \and
           Edris Esmaeili Aliabadi \at
              Islamic Azad University, Faculty of Electrical, Computer and IT Engineering, Qazvin, Iran. \\
              \email{edriss.e2006@gmail.com}
           \and
           Reza Yousefnezhad \at
              Amirkabir University of Technology, Faculty of Marine Technology, Tehran, Iran. \\
              \email{yousefnezhad@aut.ac.ir}
}

\date{Accepted: \today}
\maketitle

\begin{abstract}
From the dawn of civilization, people have used folktales and stories to share information and knowledge. After the invention of printing in the 15th century, technology provided helpful yet complicated utilities to exchange ideas. In the present computerized world, the art of storytelling is becoming more influential through the unprecedented multimedia capabilities of computers. In this article, we introduce a state-of-the-art presentation software by which academicians can present nonlinear topics efficiently and sharpen their storytelling skills. We show how the proposed software can improve the scientific presentation style. We conducted a survey to measure the attractiveness of proposed utility among other alternatives. Results show that academicians prefer the proposed platform to others.
\end{abstract}

\section{Introduction}
\label{intro}
A well-told story can be memorized and recalled quickly. People tend to learn better when the data is transformed into story, and this is the reason societies pass on their values by means of stories to the next generations. As is expressed by \citet{ref1}, a story is worth a thousand pictures since an image can talk about a single moment but a story can express the sequence of events. Therefore, developing storytelling skills has a great value. \citet{ref3} claims that storytelling skills can persuade listeners to feel more involved.

Advantages of storytelling induce researchers to exploit storytelling techniques for presentation and educational purposes. For instance, \citet{ref2} examine the performance of an interactive storytelling system for a public archaeology heritage presentation in Belgium. \citet{ref5} create a novel software architecture that couples 3D representation and storytelling for creating engaging linear narrations that can be shared on the web.

The process of information visualization can help us to provide meaningful information for viewer. However, visualization problems can become challenging due to the complexities such as extensive data volumes \citep{ref7}. \citet{ref4} address the necessity of solving high complexities with visualization problem to relieve the intrinsic limitations of human cognitive capacity and information processing ability. They suggest applying storytelling in the field of information visualization can lead to better information presentation.

Presenting scientific papers requires different qualifications than presenting general topics \citep{davis2005scientific}. In this study, we introduce new presentation software by which academicians can augment storytelling skills and present nonlinear topics efficiently. Afterward, our proposed software will be called \emph{Academic Presenter}.

The remainder of the manuscript is organized as follows. Section \ref{literature} reviews the history of available software products and introduces the proposed software. Section \ref{problems} addresses issues related to a scientific presentation that have not previously been completely solved. Section \ref{problems} also suggests our solution for each problem. Section \ref{casestudy} studies the attractiveness of the proposed software among common presentation utilities. Finally, Section \ref{conclusion} concludes.

\section{Related Work}
\label{literature}
In this section, we begin with the history of current presentation utilities and discuss the associated advantages and disadvantages of each style. Following this, we introduce the proposed software solution that creates a new paradigm in modeling visual contents by combining previous methods.

\subsection{Conventional Presentation Utilities}
From the early stages, multimedia capabilities of computers exhibited a suitability for demanding tasks such as presentation \citep{keckler2011gpus}. Initially, \emph{Presenter} was released by \emph{Forethought} \citep{ref8} and in 1987 it was renamed \emph{PowerPoint}. \emph{Microsoft} embedded \emph{PowerPoint} in the \emph{Office} suite in 1990. \emph{PowerPoint} has been designed to create linear presentations through slides. Because of high accessibility, it gained acceptance in academia \citep{pippert1999multiple}. \citet{ref11} show the positive effect of creating presentations with \emph{PowerPoint} on students' grades. However, \citet{susskind2005powerpoint} claims that \emph{PowerPoint} won't affect academic performance but enhance students' attitudes and self-efficacy about the course.

Gradually, high accessibility and linearity create issues, especially in universities \citep{tufte2003powerpoint}. Speakers create slides merely to present rather than focusing on their messages. Using slide-based presentation software together with students' lack of experience deteriorates students' organization skills. Also, the linearity of slide-based software products forces the presenter to simplify sophisticated subjects to a set of bullet items which is misleading for decision-making \citep{tufte2006cognitive}. Moreover, a linear presentation is not suitable to illustrate the complexity of an issue; nonetheless, \citet{ref9} tried to find a solution for this issue by using a directed graph structure approach.

Another movement in computer graphics started concurrently with slideware. In November 1996, \emph{Macromedia} released the first version of \emph{Flash}. \emph{Flash} is a canvas-based presentation tool that supports vector-based animation. Canvas is like an infinite and borderless workspace in which building blocks form a presentation. Unlike slide-based technology, canvas-based technology offers enough flexibility to create nonlinear presentations\footnote{ A nonlinear presentation is a presentation style in which user defines paths for illustrating the relationship among concepts by zooming, panning, and rotating screen animations \citep{good2002zoomable, bean2012presentation}}. Yet, it was difficult to create a presentation with \emph{Flash} since it required programming skill. Nowadays, new companies such as \emph{Prezi} \citep{Prezi} are trying to simplify canvas-technology for building presentations; however, this simplification may confine flexibility. \emph{Prezi} demonstrates positive results in classrooms \citep[e.g.,][]{brock2013tale, anderson2013diversifying, vspernjak2014prezi}.

Table \ref{tab:smackdown} categorizes available presentation software with respect to employed technologies and price. The first and second columns indicate whether the product is canvas-based or slide-based, respectively. The third and fourth columns determine the availability of the corresponding product as a web application (online) or conventional software (offline). Finally, the last column shows which one is free.

\begin{table}[htbp]
  \centering
  \caption{Comparison among available products}
  \resizebox{0.5\textwidth}{!}{%
    \begin{tabular}{lccccc}
    \hline
    Name  & Canvas & Slide  & Online & Offline & \multicolumn{1}{c}{Free}      \\
    \hline
    Adobe Flash \protect\footnotemark & \checkmark     &       & \checkmark     & \checkmark     &  \\
    MS PowerPoint &       & \checkmark     & \checkmark     & \checkmark     &  \\
    Prezi \protect\footnotemark & \checkmark     &       & \checkmark     & \checkmark     &  \\
    Keynote \protect\footnotemark &       & \checkmark     &       & \checkmark     &  \\
    Google Slides \protect\footnotemark &       & \checkmark     & \checkmark     &       & \multicolumn{1}{c}{\checkmark} \\
    PowToon \protect\footnotemark &       & \checkmark     & \checkmark     &       &  \\
    Academic~Presenter \protect\footnotemark & \checkmark     & \checkmark     &    $*$   & \checkmark     & \multicolumn{1}{c}{\checkmark} \\
    SlideDog \protect\footnotemark & \checkmark     & \checkmark     & \checkmark     &       &  \\
    SlideShare \protect\footnotemark &       & \checkmark     & \checkmark     &       & \multicolumn{1}{c}{\checkmark} \\
    \hline
    \multicolumn{6}{l}{$*$ \emph{Academic Presenter} supports online presentations of the designed} \\
    \multicolumn{6}{l}{projects on the offline program.} \\
    \hline
    \end{tabular}%
    }
  \label{tab:smackdown}%
\end{table}%
\addtocounter{footnote}{0}
\footnotetext{\texttt{http://www.slideshare.net}}
\addtocounter{footnote}{-1}
\footnotetext{\texttt{http://slidedog.com}}
\addtocounter{footnote}{-1}
\footnotetext{\texttt{http://www.apresenter.com}}
\addtocounter{footnote}{-1}
\footnotetext{\texttt{http://www.powtoon.com}}
\addtocounter{footnote}{-1}
\footnotetext{\texttt{https://www.google.com/slides/about}}
\addtocounter{footnote}{-1}
\footnotetext{\texttt{https://www.apple.com/mac/keynote}}
\addtocounter{footnote}{-1}
\footnotetext{\texttt{http://www.prezi.com}}
\addtocounter{footnote}{-1}
\footnotetext{\texttt{https://get.adobe.com/flashplayer}}

As one can see, only two presentation tools offer both canvas-based and slide-based technologies simultaneously. Additionally, the table imply that canvas-technology is less popular than the others, although among all presentation utilities, \emph{Adobe Flash} and \emph{Prezi} are known as revolutionary products. In this table, we added \emph{Academic Presenter} as well.

\subsection{Proposed Software Solution}
\begin{figure*}[htbp]
\centering
\includegraphics[width=1\textwidth,natwidth=2723,natheight=760]{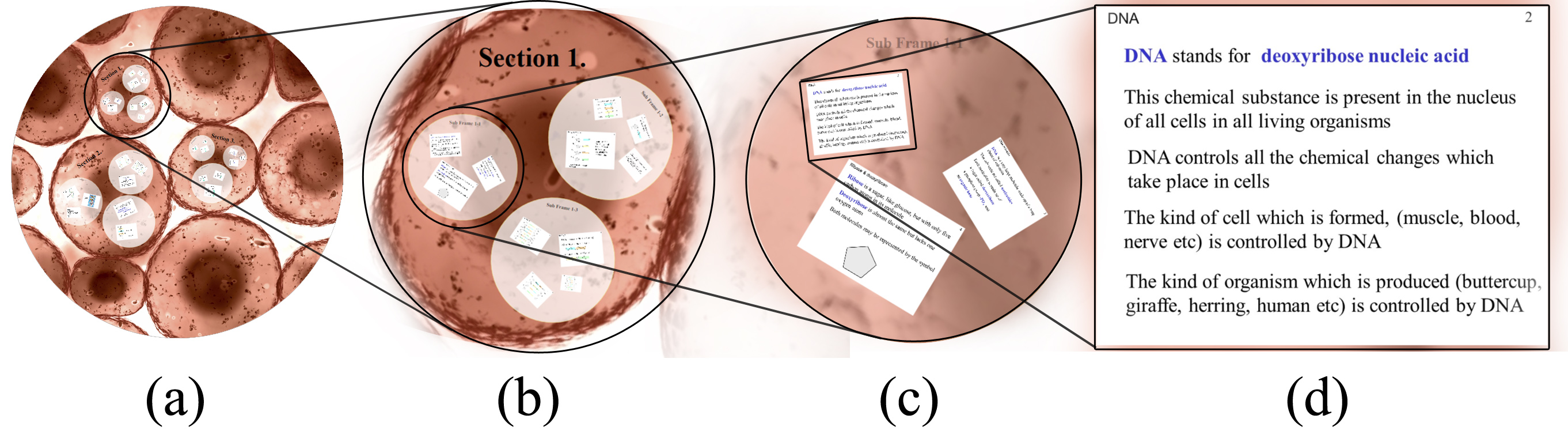}
\caption{Switching from a nonlinear flow (a, b, and c) to a linear flow at (d)}
\label{Fig-Switch}
\end{figure*}

\emph{Academic Presenter} combines the potency of slide-based presentation software products with canvas-based \footnote{ We refer the interested readers to watch \texttt{https://youtu.be/rMG8-wzCaD8} for more details.}. Users can switch between two common presentation trends based on the level of details; for introducing general topics, they can employ a nonlinear flow and switch to a conventional linear presentation for exhibiting details. Figure \ref{Fig-Switch} depicts a sample in which we used both nonlinear and linear flows. From (a) to (b) and then from (b) to (c), a user can zoom, pan, and rotate by using mouse or touch-screen. However at (d), a linear flow can carry the talk to the next topic where the user may switch to a nonlinear flow again. Thanks to the vector-based canvas of \emph{Academic Presenter}, zooming into a particular region will not affect contents' quality. By taking the advantage of proposed framework, the users can combine even mind-map diagrams and conventional slides.

\citet{chou2015prezi} investigate the effectiveness of various digital presentation tools (more specifically \emph{PowerPoint} and \emph{Prezi}) on students' learning performance. Their results show that \emph{Prezi} is a more efficient instructional medium for knowledge acquisition compared with traditional instruction; however, \emph{PowerPoint} demonstrated instructional effectiveness on only the long-term learning retention of the students compared with traditional instruction. Hence; combining the power of slideware (such as \emph{PowerPoint}) and a canvas-based product (such as \emph{Prezi}) can enhance the effectiveness of current digital presentation tools in universities. Although Table \ref{tab:smackdown} indicates that \emph{SlideDog} is also offering both presentation technologies, the user has to create \emph{PowerPoint} and \emph{Prezi} projects separately in the mentioned tools.

Our proposed software also enables users to build an engaging presentation by combining different types of audio visual contents: including image, audio, video, vector-based shape, PDF document, LaTeX code, and handwriting.

\begin{figure}[h!]
\centering
\includegraphics[width=0.5\textwidth,natwidth=1000,natheight=505]{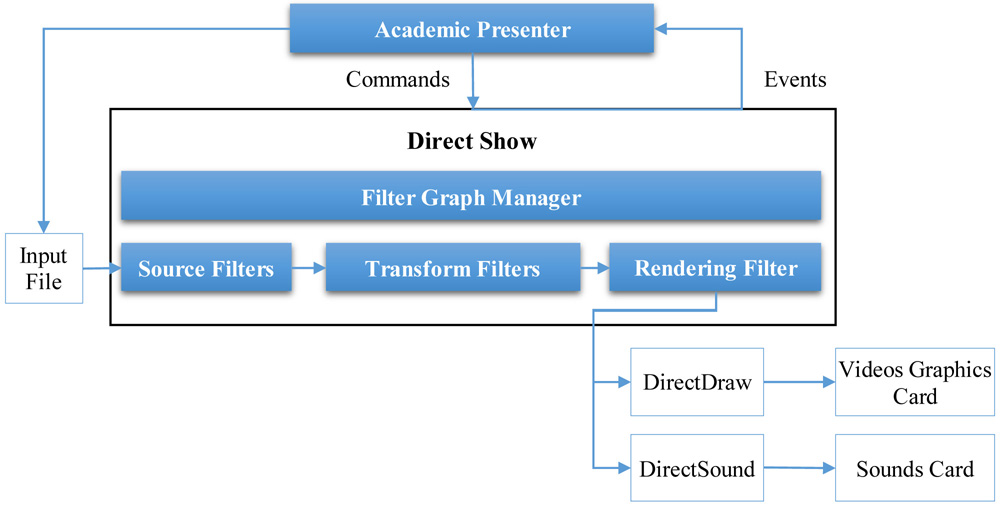}
\caption{Playing a video with DirectShow}
\label{Fig-Video}
\end{figure}

Because \emph{Academic Presenter} harnesses the power of a video graphics card without an intermediary, it is faster. Figure \ref{Fig-Video} shows the interactions among the video graphics card and application to play a video. The bottom line is that \emph{Academic Presenter} is free software, which makes it an interesting option for students on a tight-budget.

In the following sections, we focus on the application of storytelling techniques on a common scientific presentation.

\section{Application of storytelling techniques on a Scientific Presentation}
\label{problems}
In the first subsection, we propose applying mind map diagrams for presenting a typical literature review, and we explain how \emph{Academic Presenter} can help academicians with this. Next, we demonstrate how using animation and storytelling techniques can assist infographics to be more clear and informative. Finally, the effect of \emph{Academic Presenter}'s whiteboard animation on teaching quality will be discussed.

\subsection{Literature Review with Mind-map}
In any scientific presentation, researchers have to review and discuss published information. Literature review both summarizes and synthesis important information. Unfortunately, common methods to deal with literature review are as follows:
\begin{itemize}
  \item Listing the most relevant papers as bullet point items.
  \item Organizing published information inside tables and comparing them with respect to some criteria.
\end{itemize}
Indeed, these ways of organizing information are not mind-friendly since listeners have to digest and categorize information simultaneously. However, the presentation time is not enough for both thinking deeply and listening carefully. Vector-based canvas of \emph{Academic Presenter} offers another way of organizing information; using mind-map diagrams. Mind mapping has been defined as ``visual, nonlinear representations of ideas and their relationships" \citep{ref14}. Mind-map is also considered as a powerful diagramming tool that plays a significant role in collaborative or group storytelling \citep{nakamura2010zuzie}. \citet{liu2011enhanced} demonstrate the benefits of mind mapping (concept mapping) on students storytelling skills.

\begin{figure*}[h!]
\centering
\includegraphics[width=1\textwidth,natwidth=1500,natheight=958]{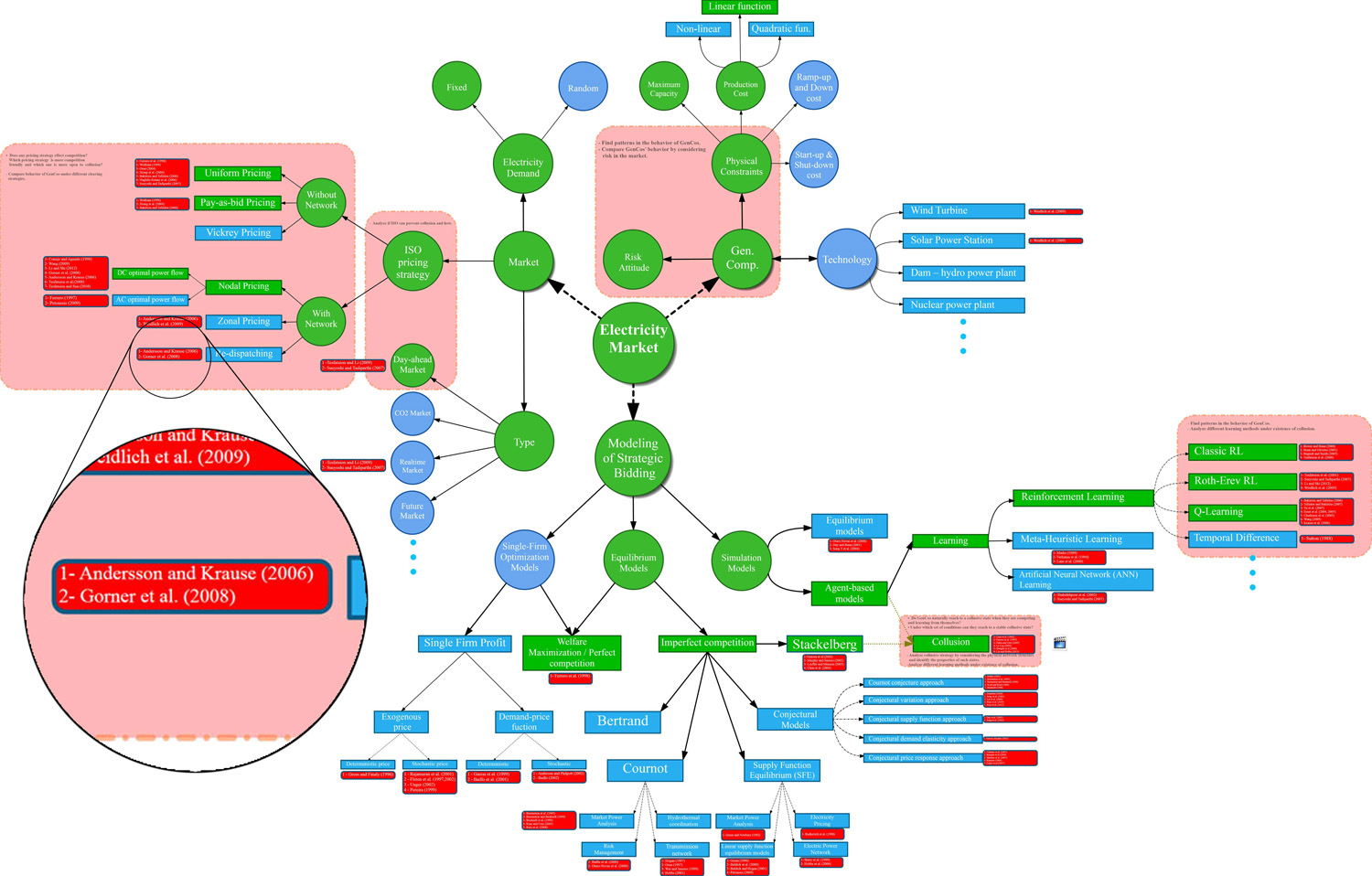}
\caption{Using mind-map to review the literature}
\label{Fig1}
\end{figure*}

By using mind-map, viewers can categorize subjects and find their relationship with the main topic. For example, Figure \ref{Fig1} depicts the literature review of a deregulated electricity market using a mind-map diagram. From the central topic toward each branch, more details are added to the parent nodes; thus, doing this provides classification rule to categorize subjects. Each branch ends with a red node containing studies similar to the attached branch. This categorization method is easier to memorize and recall \citep{ref15}. Moving from one branch to another, a presenter begins by discussing general topics and finishes with more technical information; therefore, viewers might be less likely to lose concentration as a result of listening to details for a long duration. As mentioned, \emph{Academic Presenter} supports both slide-based and canvas-based technologies; therefore, a presenter can switch to slide-mode to explain linear topics inside each node. Interested readers will be invited to watch ``Why \emph{Academic Presenter}? (Part 1 - Literature Review)"\footnote{\texttt{https://youtu.be/LUWr8pqJjzg}} for more details.

\subsection{Animated Infographics}
Information graphics (or infographics) is an innovative medium to visualize data clearly and in an engaging manner. Infographics are enchanting storytelling tools for transforming data into knowledge, as they capture a reader's attention by utilizing principles of graphic design. These characteristics assist infographics to be highly popular for transferring data to diverse audiences \citep{bateman2010useful, borkin2013makes}.

However, packing all data and information in a single image can cause a sense of confusion since people may not see the patterns clearly. \citet{harrison2015infographic} examine the impact of color and complexity on impression level of audiences and conclude that participants reacted differently to infographics due to the difference in age, educational background, and gender.

\begin{figure*}[h!]
\centering
\includegraphics[width=1\textwidth,natwidth=1500,natheight=958]{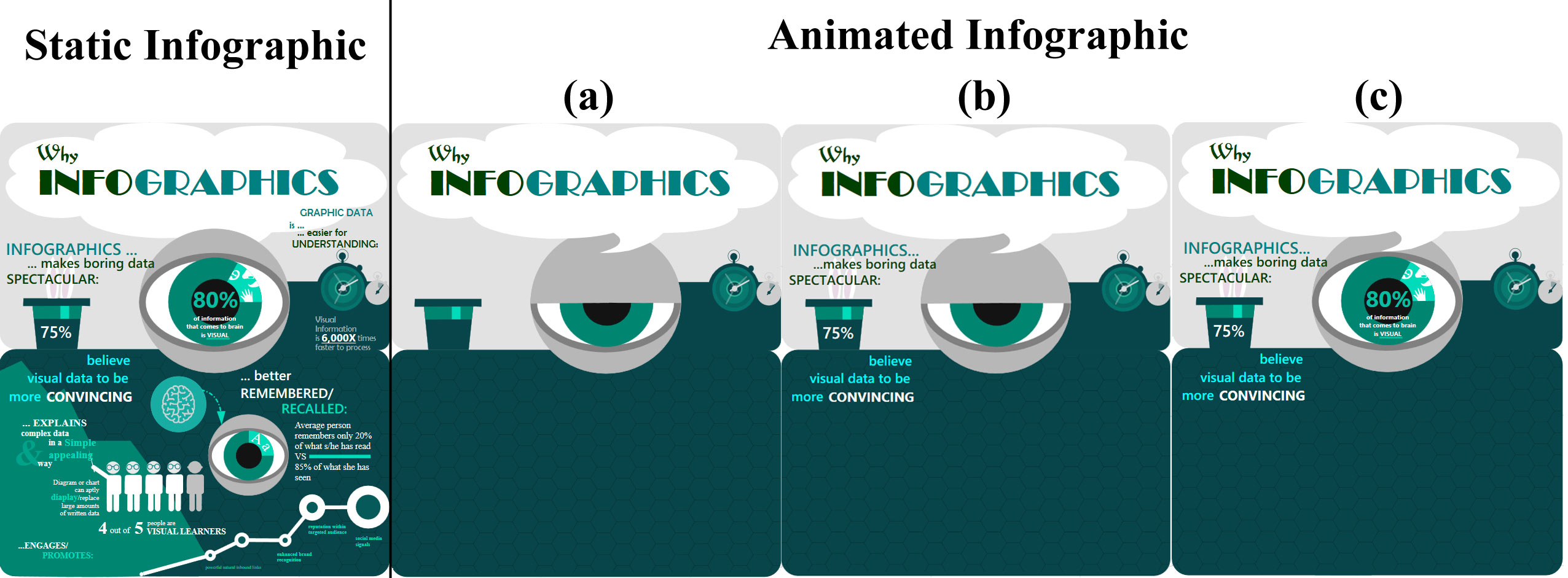}
\caption{Animating static infographics with keyframes and infinite canvas}
\label{Fig:info}
\end{figure*}

A solution to this problem is using timeline animation instead of a single image. Therefore, viewers are gradually becoming familiar with the presented data.  The combination of keyframes and infinite canvas in Academic Presenter help designers to prioritize different sections of infographics and add animation to static infographics. Figure \ref{Fig:info} displays a sample in where static infographic is converted to an animated one. In Figure \ref{Fig:info}, the leftmost image is static but the right panel is showing the development of the story with time. Note that designers can zoom and pan in each keyframe to recommend a viewport to audiences.

\subsection{Effect of Handwriting}
Although typing by computer is easier than writing by hand, there remain many debates about the constructive effects of writing by hand on learning \citep{ref13}. \citet{ref12} explains why writing by hand can assist learning. There is anecdotal evidence that dynamic sketches together with narration may be more efficient for delivering information than traditional
presentations \citep[e.g.,][]{dean2006beyond, roam2009back}. Consequently, researchers invent new teaching aids compatible with this storytelling technique. For instance, \citet{lee2013sketchstory} propose a new narrative visualization (specifically whiteboard animation) device that uses pen and touch interactions to leverage the narrative storytelling attributes. Results confirm that the audience is more engaged by presentations that done with offered tool than \emph{PowerPoint}. Besides, writing by hand allows more flexibility to the writer, especially in abstract courses such as mathematics. Nowadays, many educational websites are using whiteboard animation technique to teach various topics \citep[e.g.,][]{ASAPScience, RSAAnimate}.

\begin{figure}[h!]
\centering
\includegraphics[width=0.5\textwidth,natwidth=407,natheight=169]{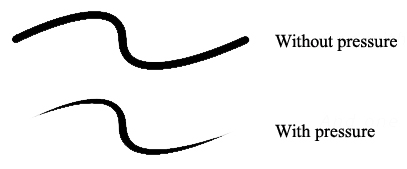}
\caption{Effect of ignoring pressure on a stroke}
\label{Fig3}
\end{figure}

\begin{figure}[h!]
\centering
\includegraphics[width=0.5\textwidth,natwidth=1000,natheight=734]{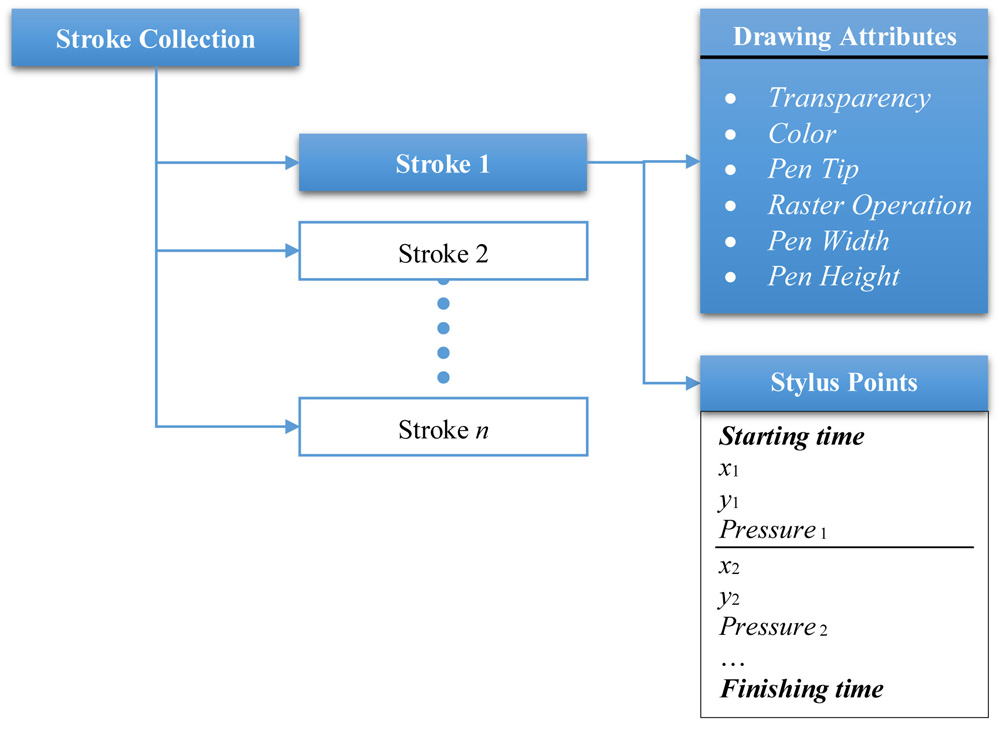}
\caption{Data structure of digital ink class}
\label{Fig-Pen}
\end{figure}

In spite of progress in teaching instruments, many professors still prefer to teach by writing on a board. However, by looking at the entire academic career of a professor, one might infer that s/he often teaches almost the same materials each semester to different groups of students. We suggest employing digitizer to utilize the advantages of writing by hand yet alleviate the repetition issue. Nowadays, digitizers are becoming an indispensable part of any computer. Users can record their hand movements on screen by using digitizers. The information which can be retrieved from digitizers is as follows: 2D-position, pressure level, starting time, finishing time, and color. Each time the user draws a line (stroke) on screen, the digitizer records the position of the digitizer's tip on screen and pressure level. The pressure sensitivity of all digitizers is not the same, but even low-quality digitizers can sense the pressure accurately enough to emulate the movement. Figure \ref{Fig3} shows effect of neglecting pressure on a stroke. Figure \ref{Fig-Pen} illustrates the employed data structure. The stroke collection consists of strokes and each stroke corresponds to one curve on the canvas.

\begin{figure*}[h!]
\centering
\includegraphics[width=0.7\textwidth,natwidth=1600,natheight=858]{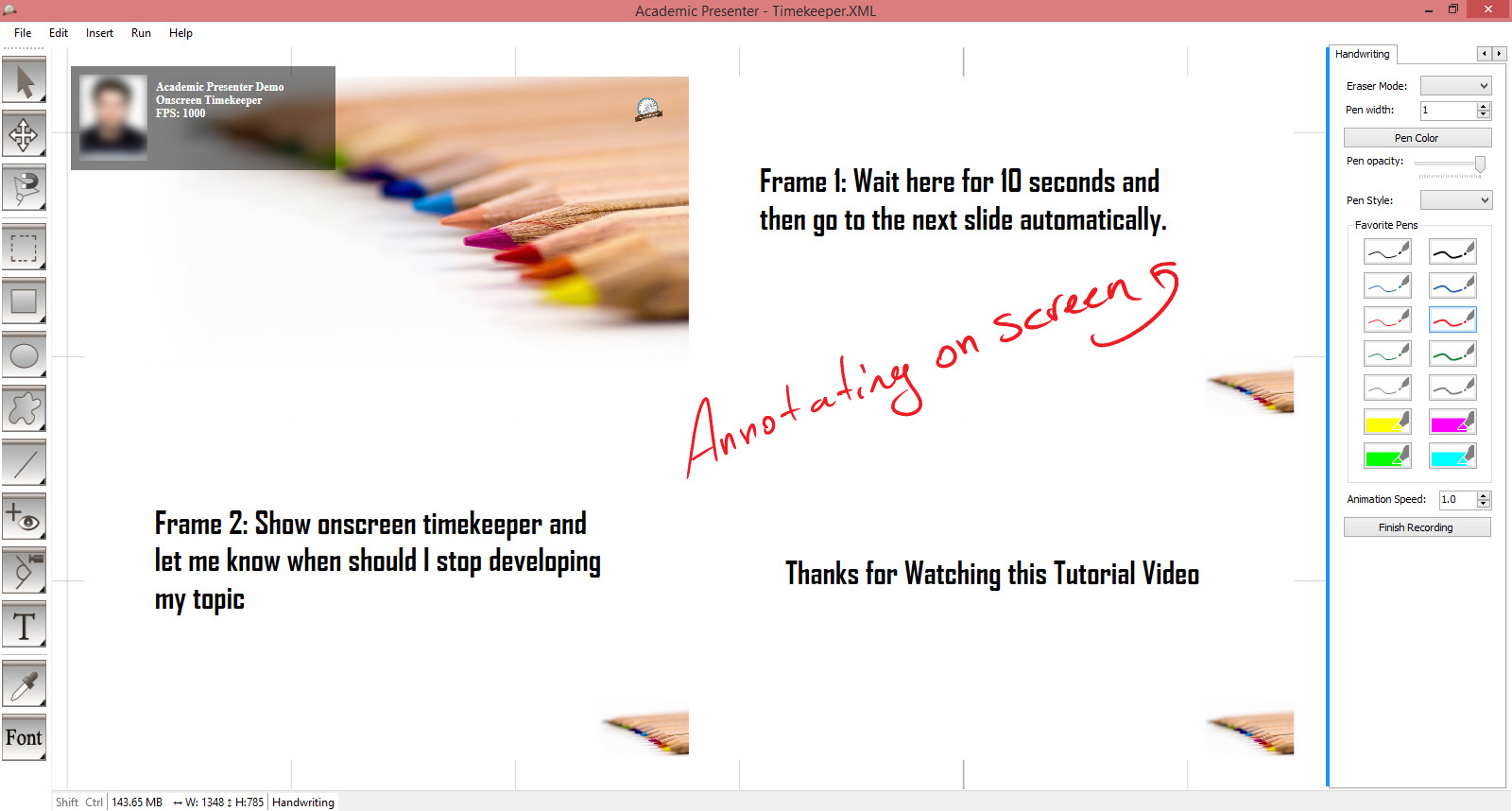}
\caption{Handwriting toolbox in \emph{Academic Presenter}}
\label{Fig_HW}
\end{figure*}

\emph{Academic Presenter} utilizes a digitizer in presentation, not only to annotate on screen but also to replay the handwriting wherever is necessary. In toolbox, a handful of different pens and highlighters is available at users' fingertips. Figure \ref{Fig_HW} displays the handwriting toolbox. Every movement is editable and precise. Also, user can increase animation speed to save presentation time. We redirect an interested reader to watch ``Why \emph{Academic Presenter}? (Part 5 - Handwriting)" \footnote{\texttt{https://youtu.be/U-oNFjBtzfE}} for more details.

\section{Analyzing the Attractiveness of Academic Presenter}
\label{casestudy}
\emph{Academic Presenter} is designed for academic environments since presentation has educational and inspirational nature. Potential users are students of universities, teachers in high school and professors. Because our focus is to solve presentation problems related to academic environments, we tried to find flaws in current tools which affect the presentations the most. To analyze the future position of \emph{Academic Presenter} in academia, we exploit Analytic Hierarchy Process (AHP) method \citep{saaty1988analytic}; therefore, we define four criteria: \emph{Price}, \emph{Number of users}, \emph{Simplicity}, and \emph{Applicability in academia}. We collect quantitative information such as price and number of users from the websites and documents; however for qualitative criterion such as simplicity, we asked from experts in the field of presentation. We select the most significant competitors. The alternatives are listed as follows:

\begin{itemize}
  \item \textbf{Office 365} (including \emph{PowerPoint}) is the brand name adopted by Microsoft for a collection of software plus services subscriptions that provides web-based productivity software and services to its subscribers.
  \item \textbf{Prezi} is a cloud-based storytelling tool for presenting ideas on a virtual canvas. The product employs a zooming user interface, which allows users to zoom in and out of their visual contents, and enables users to navigate through information within a 2.5D space on the Z-axis.
  \item \textbf{SlideShare} is a web-based slide hosting service. Users can upload \emph{PowerPoint}, PDF, and Keynote files privately or publicly. Slide decks can then be viewed on the site itself, on hand held devices or embedded on other websites. \emph{SlideShare} is considered to be similar to YouTube, but for slide shows.
  \item \textbf{PowToon} is a cloud-based for creating animated presentations and animated explainer videos.
  \item \textbf{emaze} is an online presentation platform built on html5 technology. Users can create, manage and share their presentations through their cloud-based system. It offers 3D animations and video backgrounds.
\end{itemize}
The retrieved information from competitors are displayed in Table \ref{tab:info}.
\begin{table*}[htbp]
  \centering
  \caption{Retrieved information from competitors }
  %\resizebox{1\textwidth}{!}{%
    \begin{tabular}{lcccc}
    \hline
    Presentation   & Price     & Number of users & \multirow{2}{*}{Simplicity} & Applicability \\
    tools          & (\$/year) & (millions)      &                             & in academia   \\
    \hline
    Academic Presenter & 0     & 0.022 & 0.5   & 0.75 \\
    Office 365         & 79.99 & 15.2  & 0.5   & 0.88 \\
    Prezi              & 159   & 40    & 0.8   & 0.58 \\
    SlideShare         & 228   & 70    & 1     & 0.25 \\
    PowToon            & 228   & 6     & 0.5   & 0.50 \\
    emaze              & 178.92& 0.011 & 1     & 0.58 \\
    \hline
    \end{tabular}%
  %  }
  \label{tab:info}%
\end{table*}%

As one can perceive from Table \ref{tab:applicability}, applicability in academia is calculated based on availability of essential features that may help students and professors during their presentations. Also, there are some features with half the unit value for some alternatives which means mentioned feature is not provided at a satisfactory level.
\begin{table*}[htbp]
  \centering
  \caption{Calculating applicability in academia based on important features}
  %\resizebox{1\textwidth}{!}{%
    \begin{tabular}{lcccccc}
    \hline
    Applicability & Academic  & Office & \multirow{2}{*}{Prezi} & \multirow{2}{*}{Slideshare} & \multirow{2}{*}{PowToon} & \multirow{2}{*}{emaze} \\
    in academia   & Presenter & 365    &       &                                &         &       \\
    \hline
    Supporting Images & 1     & 1     & 1     & 0     & 1     & 1 \\
    Supporting Sounds & 1     & 1     & 1     & 0     & 1     & 1 \\
    Supporting Videos & 1     & 1     & 1     & 0     & 1     & 1 \\
    Formula and Latex & 1     & 1     & 0     & 0     & 0     & 0 \\
    Online Presentation & 0.5   & 1     & 1     & 1     & 1     & 1 \\
    Offline Presentation & 1     & 1     & 1     & 0     & 0     & 0 \\
    Nonlinear Presentation & 1     & 0     & 1     & 0     & 0     & 0 \\
    Linear Presentation & 1     & 1     & 0     & 1     & 1     & 1 \\
    Annotation & 1     & 0.5   & 0     & 0     & 0     & 0 \\
    Supporting Second Screen & 0     & 1     & 0     & 0     & 0     & 0 \\
    Charts & 0     & 1     & 0     & 0     & 0     & 1 \\
    Running on different OSs & 0.5   & 1     & 1     & 1     & 1     & 1 \\
    \hline
    Weight & 0.75  & 0.88  & 0.58  & 0.25  & 0.50  & 0.58 \\
    \hline
    \end{tabular}%
%    }
  \label{tab:applicability}%
\end{table*}%

\begin{figure}[h!]
\centering
\includegraphics[width=0.5\textwidth,natwidth=1075,natheight=667]{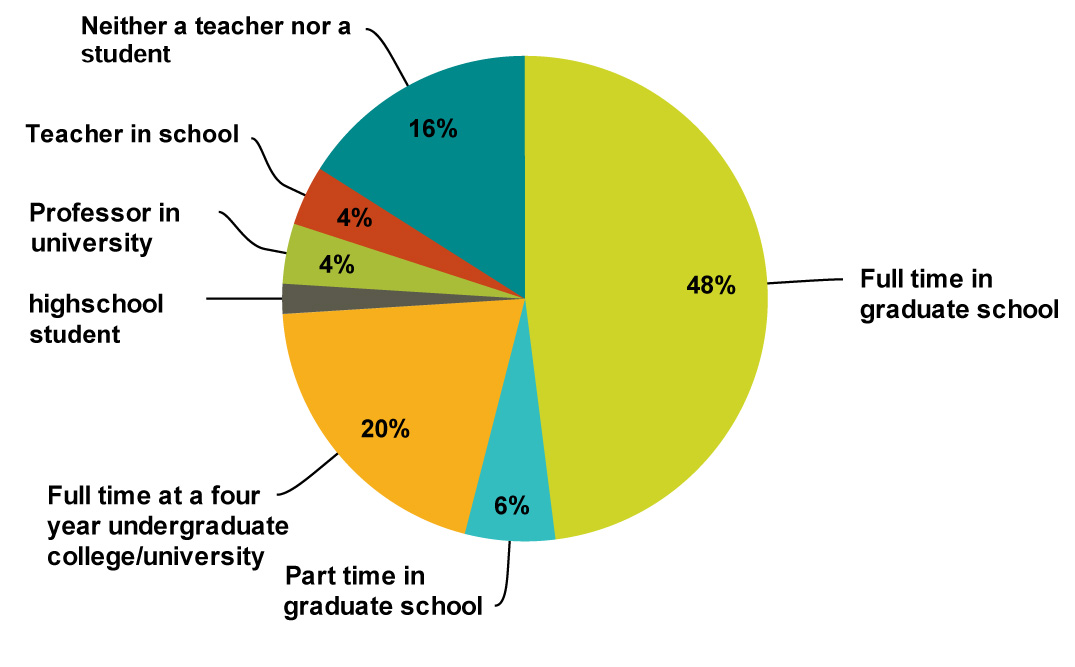}
\caption{The detailed information of the attendees in the survey}
\label{Fig-Chart}
\end{figure}

Furthermore, we invite users to judge about the importance of each criterion. A group of 50 people have attended in a questionnaire. The composition of the attendees are as follows: graduate students 54\%, undergraduate students 20\%, and instructors 8\%. Figure \ref{Fig-Chart} delineates the detailed information of the participants on a pie chart.

The resulted judgements are reported in Table \ref{tab:judgement}. Based upon pairwise comparisons, applicability is the most influential factor. The inconsistency of judgement matrix is 1\% which is in acceptable range.
\begin{table*}[htbp]
  \centering
  \caption{The relative importance of each criterion}
  %\resizebox{1\textwidth}{!}{%
    \begin{tabular}{l|cccc|c}
    \hline
    Criterion& \multicolumn{1}{c}{{Price}} & \multicolumn{1}{c}{{Num. of users}} & \multicolumn{1}{c}{{Applicability}} & \multicolumn{1}{c}{{Simplicity}} &  {Weight}\\
    \hline
    {Price}         & 1     & 1.223 & 0.820    &  0.888  &  0.241 \\
    {Num. of users} & 0.818 & 1     & 0.670    &  0.670  &  0.193 \\
    {Applicability} & 1.220 & 1.492 & 1        &  1.084  &  0.294 \\
    {Simplicity}    & 1.126 & 1.377 & 0.923    &  1      &  0.271 \\
    \hline
    \end{tabular}%
  %  }
  \label{tab:judgement}%
\end{table*}%

\begin{figure}[h!]
\centering
\includegraphics[width=0.5\textwidth,natwidth=387,natheight=131]{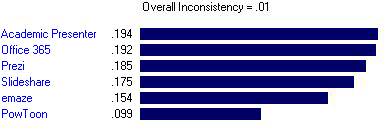}
\caption{Alternatives' ranking based on AHP method}
\label{Fig-Alt}
\end{figure}

AHP estimates \emph{Academic Presenter}'s position among competitors regarding retrieved information and pairwise judgements (see Figure \ref{Fig-Alt}).

\begin{figure*}[h!]
\centering
\includegraphics[width=0.75\textwidth,natwidth=1920,natheight=878]{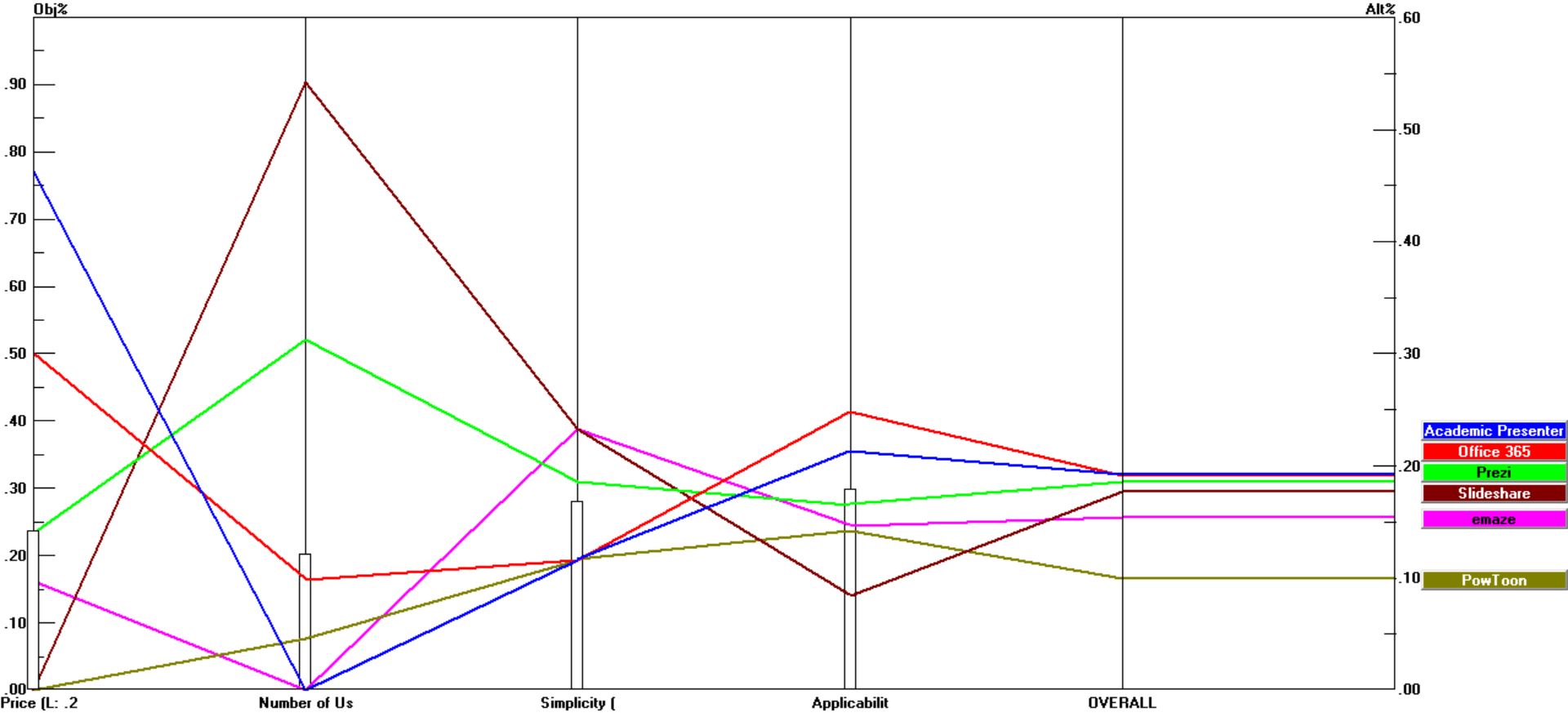}
\caption{AHP ranking of alternatives considering for each criterion}
\label{Fig-AHP}
\end{figure*}

As one can see in Figure \ref{Fig-AHP}, \emph{Academic Presenter} had better off in term of price. Sensitivity analysis of our result is showing that \emph{Academic Presenter}'s rank is relatively stable on simplicity and applicability. Although \emph{Academic Presenter} is showing a promising rank among other alternatives, yet the difference between \emph{Prezi}, \emph{Office 365} and \emph{Academic Presenter} is negligible.

\section{Conclusion}
\label{conclusion}
Presenting scientific papers need different requirements than presenting general topics. Most of available software solutions are adjusted to meet business presentations' demands. However, presenting a nonlinear scientific subject is beyond their capabilities. In this paper, we presented a new presentation software that facilitates delivering nonlinear topics. Our free presentation software enables users to enhance their storytelling skill. Users can switch between two common presentation trends based on the level of details; for introducing general topics, they can employ a nonlinear flow and switch to a conventional linear presentation for exhibiting details. Also, we introduce new components in the presented software solution that may help academicians to teach abstract courses more efficiently. Finally, a survey is conducted by asking eligible attendees to prioritize different aspects of a presentation utility. We exploit Analytic Hierarchy Process method to analyze the expected rank of proposed tool among popular alternatives. The results are indicating that the proposed software is more attractive than current software solutions.

Although the proposed utility is the combination of slide-based and canvas-based products and researchers investigated on each technology separately, assessing the effectiveness of proposed tool on the knowledge acquisition of students is a valuable future work.

\begin{acknowledgements}
Authors of this article would like to express their very great appreciation to Daniel Lee Calvey for his valuable and constructive suggestions. Also, the authors would like to thank all those who cooperated and dedicated their valuable times in the survey.
\end{acknowledgements}
\section*{Compliance with ethical standards}
\textbf{Conflict of interest } This study was self-funded and no conflict of interest exists.

\bibliographystyle{spbasic}
\bibliography{Ref}

\begin{thebibliography}{37}
\providecommand{\natexlab}[1]{#1}
\providecommand{\url}[1]{{#1}}
\providecommand{\urlprefix}{URL }
\expandafter\ifx\csname urlstyle\endcsname\relax
  \providecommand{\doi}[1]{DOI~\discretionary{}{}{}#1}\else
  \providecommand{\doi}{DOI~\discretionary{}{}{}\begingroup
  \urlstyle{rm}\Url}\fi
\providecommand{\eprint}[2][]{\url{#2}}

\bibitem[{Anderson et~al.(2013)Anderson, Whitefield, Virtanen, Myll{\"a}rniemi,
  and Wallander}]{anderson2013diversifying}
Anderson G, Whitefield T, Virtanen P, Myll{\"a}rniemi J, Wallander H (2013)
  Diversifying higher education: facilitating different ways of learning.
  Campus-Wide Information Systems 30(3):201--211

\bibitem[{Royal Society for the encouragement~of Arts and
  Commerce(2015)}]{RSAAnimate}
Royal Society for the encouragement~of Arts M, Commerce (2015) {RSA} -
  {A}nimate. Retrived from http://thersa.org/events/rsaanimate,
  \urlprefix\url{http://thersa.org/events/rsaanimate}

\bibitem[{Baccarani et~al.(2015)Baccarani, Bonfanti, and Elving}]{ref3}
Baccarani C, Bonfanti A, Elving WL (2015) Effective public speaking: a
  conceptual framework in the corporate-communication field. Corporate
  Communications: An International Journal 20(3)

\bibitem[{Bai et~al.(2009)Bai, White, and Sundaram}]{ref7}
Bai X, White D, Sundaram D (2009) Visual intelligence density: definition,
  measurement, and implementation. In: Proceedings of the 10th international
  Conference NZ Chapter of the ACM's Special interest Group on Human-Computer
  interaction, ACM, pp 93--100

\bibitem[{Bai et~al.(2015)Bai, White, and Sundaram}]{ref4}
Bai X, White D, Sundaram D (2015) Towards a flexible user-centred visual
  presentation approach

\bibitem[{Bateman et~al.(2010)Bateman, Mandryk, Gutwin, Genest, McDine, and
  Brooks}]{bateman2010useful}
Bateman S, Mandryk RL, Gutwin C, Genest A, McDine D, Brooks C (2010) Useful
  junk?: the effects of visual embellishment on comprehension and memorability
  of charts. In: Proceedings of the {SIGCHI} Conference on Human Factors in
  Computing Systems, ACM, pp 2573--2582

\bibitem[{Bean(2012)}]{bean2012presentation}
Bean JW (2012) Presentation software supporting visual design: Displaying
  spatial relationships with a zooming user interface. In: Professional
  Communication Conference (IPCC), IEEE, pp 1--6

\bibitem[{Biktimirov and Nilson(2006)}]{ref14}
Biktimirov EN, Nilson LB (2006) Show them the money: Using mind mapping in the
  introductory finance course. Journal of Financial Education pp 72--86

\bibitem[{Borkin et~al.(2013)Borkin, Vo, Bylinskii, Isola, Sunkavalli, Oliva,
  and Pfister}]{borkin2013makes}
Borkin MA, Vo AA, Bylinskii Z, Isola P, Sunkavalli S, Oliva A, Pfister H (2013)
  What makes a visualization memorable? Transactions on Visualization and
  Computer Graphics 19(12):2306--2315

\bibitem[{Brock and Brodahl(2013)}]{brock2013tale}
Brock S, Brodahl C (2013) A tale of two cultures: Cross cultural comparison in
  learning the {Prezi} presentation software tool in the {US} and {N}orway. In:
  Proceedings of the Informing Science and Information Technology Education
  Conference, vol 2013, pp 95--119

\bibitem[{Chou et~al.(2015)Chou, Chang, and Lu}]{chou2015prezi}
Chou PN, Chang CC, Lu PF (2015) {Prezi} versus {PowerPoint}: The effects of
  varied digital presentation tools on students' learning performance.
  Computers and Education \doi{10.1016/j.compedu.2015.10.020}

\bibitem[{Davis et~al.(2012)Davis, Davis, and Dunagan}]{davis2005scientific}
Davis M, Davis KJ, Dunagan M (2012) Scientific papers and presentations.
  Academic Press

\bibitem[{Dean(2006)}]{dean2006beyond}
Dean BC (2006) Beyond screen capture: Creating effective multimedia whiteboard
  lectures on a tablet pc. In: Proc. Annual Workshop on the Impact of Pen
  Technology in Education (WIPTE)

\bibitem[{Del and Theresa(2001)}]{ref11}
Del S, Theresa F (2001) Laptop computers and multimedia and presentation
  software. Journal of Research on Technology in Education 34(1):29--37,
  \doi{10.1080/15391523.2001.10782331},
  \urlprefix\url{http://dx.doi.org/10.1080/15391523.2001.10782331}

\bibitem[{Farrand et~al.(2002)Farrand, Hussain, and Hennessy}]{ref15}
Farrand P, Hussain F, Hennessy E (2002) The efficacy of the mind map'study
  technique. Medical education 36(5):426--431

\bibitem[{Gaskins(1984)}]{ref8}
Gaskins R (1984) Sample product proposal: Presentation graphics for overhead
  projection. Retrieved on July 16, 2011

\bibitem[{Gatto and Pittarello(2014)}]{ref5}
Gatto I, Pittarello F (2014) Creating {Web3D} educational stories from
  crowdsourced annotations. Journal of Visual Languages and Computing
  25(6):808--817

\bibitem[{Gershon and Page(2001)}]{ref1}
Gershon N, Page W (2001) What storytelling can do for information
  visualization. Communications of the ACM 44(8):31--37

\bibitem[{Good and Bederson(2002)}]{good2002zoomable}
Good L, Bederson BB (2002) Zoomable user interfaces as a medium for slide show
  presentations. Information Visualization 1(1):35--49

\bibitem[{Harrison et~al.(2015)Harrison, Reinecke, and
  Chang}]{harrison2015infographic}
Harrison L, Reinecke K, Chang R (2015) Infographic aesthetics: Designing for
  the first impression. In: Proceedings of the 33rd Annual {ACM} Conference on
  Human Factors in Computing Systems, ACM, pp 1187--1190

\bibitem[{Keckler et~al.(2011)Keckler, Dally, Khailany, Garland, and
  Glasco}]{keckler2011gpus}
Keckler SW, Dally WJ, Khailany B, Garland M, Glasco D (2011) {GPU}s and the
  future of parallel computing. IEEE Micro (5):7--17

\bibitem[{Lee et~al.(2013)Lee, Kazi, and Smith}]{lee2013sketchstory}
Lee B, Kazi RH, Smith G (2013) Sketchstory: Telling more engaging stories with
  data through freeform sketching. IEEE Transactions on Visualization and
  Computer Graphics 19(12):2416--2425

\bibitem[{Liu et~al.(2011)Liu, Chen, Shih, Huang, and Liu}]{liu2011enhanced}
Liu CC, Chen HS, Shih JL, Huang GT, Liu BJ (2011) An enhanced concept map
  approach to improving children's storytelling ability. Computers and
  Education 56(3):873--884

\bibitem[{Longcamp et~al.(2005)Longcamp, Zerbato-Poudou, and Velay}]{ref13}
Longcamp M, Zerbato-Poudou MT, Velay JL (2005) The influence of writing
  practice on letter recognition in preschool children: A comparison between
  handwriting and typing. Acta Psychologica 119(1):67--79,
  \doi{http://dx.doi.org/10.1016/j.actpsy.2004.10.019},
  \urlprefix\url{http://www.sciencedirect.com/science/article/pii/S0001691804001167}

\bibitem[{Moffit and Brown(2015)}]{ASAPScience}
Moffit M, Brown G (2015) Asap{SCIENCE}. Retrived from
  http://www.youtube.com/user/AsapSCIENCE,
  \urlprefix\url{http://www.youtube.com/user/AsapSCIENCE}

\bibitem[{Nakamura et~al.(2010)Nakamura, Kobayakawa, Takami, Tsuruga, Kubota,
  Hamasaki, Nishimura, and Sunaga}]{nakamura2010zuzie}
Nakamura Y, Kobayakawa M, Takami C, Tsuruga Y, Kubota H, Hamasaki M, Nishimura
  T, Sunaga T (2010) Zuzie: Collaborative storytelling based on multiple
  compositions. In: Interactive Storytelling, Springer, pp 117--122

\bibitem[{Perron and Stearns(2011)}]{Prezi}
Perron BE, Stearns AG (2011) A review of a presentation technology: {Prezi}.
  Research on Social Work Practice 21(3):376--377,
  \doi{10.1177/1049731510390700}

\bibitem[{Pinola(2011)}]{ref12}
Pinola M (2011) Why you learn more effectively by writing than typing. Retrived
  from
  http://lifehacker.com/5738093/why-you-learn-more-effectively-by-writing-than-typing

\bibitem[{Pippert and Moore(1999)}]{pippert1999multiple}
Pippert TD, Moore HA (1999) Multiple perspectives on multimedia in the large
  lecture. Teaching Sociology pp 92--109

\bibitem[{Pletinckx et~al.(2003)Pletinckx, Silberman, and Callebaut}]{ref2}
Pletinckx D, Silberman N, Callebaut D (2003) Heritage presentation through
  interactive storytelling: a new multimedia database approach. The Journal of
  Visualization and Computer Animation 14(4):225--231

\bibitem[{Roam(2009)}]{roam2009back}
Roam D (2009) The back of the napkin (expanded edition): Solving problems and
  selling ideas with pictures. Penguin

\bibitem[{Saaty(1988)}]{saaty1988analytic}
Saaty TL (1988) What is the analytic hierarchy process? Springer

\bibitem[{{\v{S}}pernjak(2014)}]{vspernjak2014prezi}
{\v{S}}pernjak A (2014) Is prezi more usefulness education tool than
  powerpoint? In: International Conference on Education in Mathematics,
  Science, and Technology, pp 406--410

\bibitem[{Spicer et~al.(2012)Spicer, Lin, Kelliher, and Sundaram}]{ref9}
Spicer R, Lin YR, Kelliher A, Sundaram H (2012) {NextSlidePlease}: Authoring
  and delivering agile multimedia presentations. {ACM} Transactions on
  Multimedia Computing, Communications, and Applications 8(4):53:1--53:20,
  \doi{10.1145/2379790.2379795},
  \urlprefix\url{http://doi.acm.org/10.1145/2379790.2379795}

\bibitem[{Susskind(2005)}]{susskind2005powerpoint}
Susskind JE (2005) {PowerPoint}'s power in the classroom: Enhancing students'
  self-efficacy and attitudes. Computers and Education 45(2):203--215

\bibitem[{Tufte(2003)}]{tufte2003powerpoint}
Tufte E (2003) {PowerPoint} is evil: Power corrupts. {PowerPoint} corrupts
  absolutely. \urlprefix\url{http://www.wired.com/2003/09/ppt2}

\bibitem[{Tufte(2006)}]{tufte2006cognitive}
Tufte E (2006) The cognitive style of {PowerPoint}: pitching out corrupts
  within

\end{thebibliography}
\end{document}